\documentclass[twocolumn,english,prl,superscriptaddress]{revtex4-1}
\usepackage[T1]{fontenc}
\usepackage[latin9]{inputenc}
\usepackage{bm}
\usepackage{graphicx}
\usepackage{natbib}
\usepackage[colorlinks,linkcolor=blue,citecolor=blue,urlcolor=blue]{hyperref}

\usepackage{dcolumn}
\usepackage{tikz}
\usepackage{amsmath,amsfonts,amssymb,natbib}

\global\long\def\ui{\mathbbm{i}}
\global\long\def\ud{\mathrm{d}}

\usepackage{pifont}
\newcommand{\cmark}{\ding{51}}
\newcommand{\xmark}{\ding{55}}
\usepackage{makecell}

\makeatletter
\usepackage{bbm}
\usepackage[bb=boondox]{mathalfa}
\makeatother

\usepackage{babel}
\begin{document}
\title{Unitary Symmetry-Protected Non-Abelian Statistics of Majorana Modes}

\author{Jian-Song Hong}
\thanks{These authors contribute equally to this work.}
\affiliation{International Center for Quantum Materials and School of Physics, Peking University, Beijing 100871, China}
\affiliation{Collaborative Innovation Center of Quantum Matter, Beijing 100871, China}
\author{Ting-Fung Jeffrey Poon}
\thanks{These authors contribute equally to this work.}
\affiliation{International Center for Quantum Materials and School of Physics, Peking University, Beijing 100871, China}
\affiliation{Collaborative Innovation Center of Quantum Matter, Beijing 100871, China}
\author{Long Zhang}
\affiliation{International Center for Quantum Materials and School of Physics, Peking University, Beijing 100871, China}
\affiliation{Collaborative Innovation Center of Quantum Matter, Beijing 100871, China}
\author{Xiong-Jun Liu}
\thanks{Correspondence addressed to: xiongjunliu@pku.edu.cn}
\affiliation{International Center for Quantum Materials and School of Physics, Peking University, Beijing 100871, China}
\affiliation{Collaborative Innovation Center of Quantum Matter, Beijing 100871, China}
\affiliation{CAS Center for Excellence in Topological Quantum Computation, University of Chinese Academy of Sciences, Beijing 100190, China}
\affiliation{Institute for Quantum Science and Engineering and Department of Physics, Southern University of Science and Technology, Shenzhen 518055, China}
%
%
\begin{abstract}
Symmetry-protected topological superconductors (TSCs) can host multiple Majorana zero modes (MZMs) at their edges or vortex cores, while whether the Majorana braiding in such systems is non-Abelian in general remains an open question.
Here we uncover in theory the unitary symmetry-protected non-Abelian statisitcs of MZMs and propose the experimental realization. We show that braiding two vortices with each hosting $N$ unitary symmetry-protected MZMs generically reduces to $N$ independent sectors, with each sector braiding two different Majorana modes. This renders the unitary symmetry-protected non-Abelian statistics.
As a concrete example, we demonstrate the proposed non-Abelian statistics in a spin-triplet TSC which hosts two MZMs at each vortex and, interestingly, can be precisely mapped to a quantum anomalous Hall insulator. Thus the unitary symmetry-protected non-Abelian statistics can be verified in the latter insulating phase, with the application to realizing various topological quantum gates being studied. Finally, we propose a novel experimental scheme to realize the present study in an optical Raman lattice. Our work opens a new route for Majorana-based topological quantum computation.
\end{abstract}

\maketitle

{\em Introduction.}--Majorana zero modes (MZMs) are self-Hermitian modes residing at the ends of one-dimensional (1D) topological superconductors (TSCs)~\cite{Kitaev2001} or the vortex cores of 2D TSCs~\cite{ReadGreen2000}. Being of an irrational quantum dimension $\sqrt{2}$~\cite{Nayak2008}, the MZMs obey non-Abelian braiding statistics~\cite{Nayak1996,Ivanov2001,DasSarma2005,Alicea2011} which can be potentially applied to topological quantum computation~\cite{Kitaev2003,Nayak2008,Aasen2016,Pachos2017}. The search for MZMs has stimulated great efforts in experiment~\cite{Mourik2012,MTDeng2012,Rokhinson2012,Shtrikman2012,JFJia2012, Marcus2013,JFJia2014,Yazdani2014,Marcus2015,Marcus2016,Molenkamp2016, Molenkamp2017,Qiao2017,HDing2018a,HDing2018b,Marcus2019, Molenkamp2019,Yazdani2019}. MZMs may appear as multiplets in TSCs when there is symmetry protection, e.g. in time-reversal (TR) invariant TSC Majorana modes come in pairs due to Kramers' theorem~\cite{SCZhang2009,TeoKane2010,Timm2010,Beenakker2011, KTLaw2012,Nagaosa2012,KaneMele2013,Berg2013,Oreg2019}. It was proposed that Majorana Kramers pairs obey non-Abelian braiding due to the protection of TR symmetry, leading to the notion of symmetry-protected non-Abelian statistics~\cite{XJLiu2014,XJLiu2016}. However, as TR symmetry is anti-unitary, even the TSC is TR invariant, the braiding, which is characterized by a unitary evolution, may dynamically break the TR symmetry~\cite{XJLiu2016,Cooper2018}. 
Such dynamical symmetry-breaking causes local mixing in Majorana Kramers pair~\cite{localmixing1,localmixing2,Knapp2020}. To achieve the symmetry-protected non-Abelian statistics of Majorana Kramers pairs then necessitates extra symmetry condition which recovers TR symmetry in the dynamical braiding process~\cite{XJLiu2016}.

In this work, we propose and establish a generic theory for non-Abelian statistics of MZMs protected by unitary symmetries. Unlike the TR symmetry, the unitary symmetry can generically protect non-Abelian statistics without suffering the dynamical symmetry breaking in the braiding process. We further demonstrate the proposed non-Abelian braiding for MZM pairs bound to vortex cores in a concrete model, which belongs to the family of spin-triplet TSCs~\cite{spintri2003,spintri2006,spintri2008,spintri2010,spintri2010b,spintri2019}. By mapping the TSC to a quantum anomalous Hall insulator (QAHI) with zero modes via particle-hole transformation, we realize the non-Abelian statistics and various quantum logic gates in the latter system. A novel scheme with experimental feasibility is proposed to for the realization 
based on the recently widely studied optical Raman lattices~\cite{ORL2014,ORL2016,ORL2019,ORL2020}, and the results are numerically verified. 

\begin{figure}
\begin{centering}
\includegraphics[width=\columnwidth]{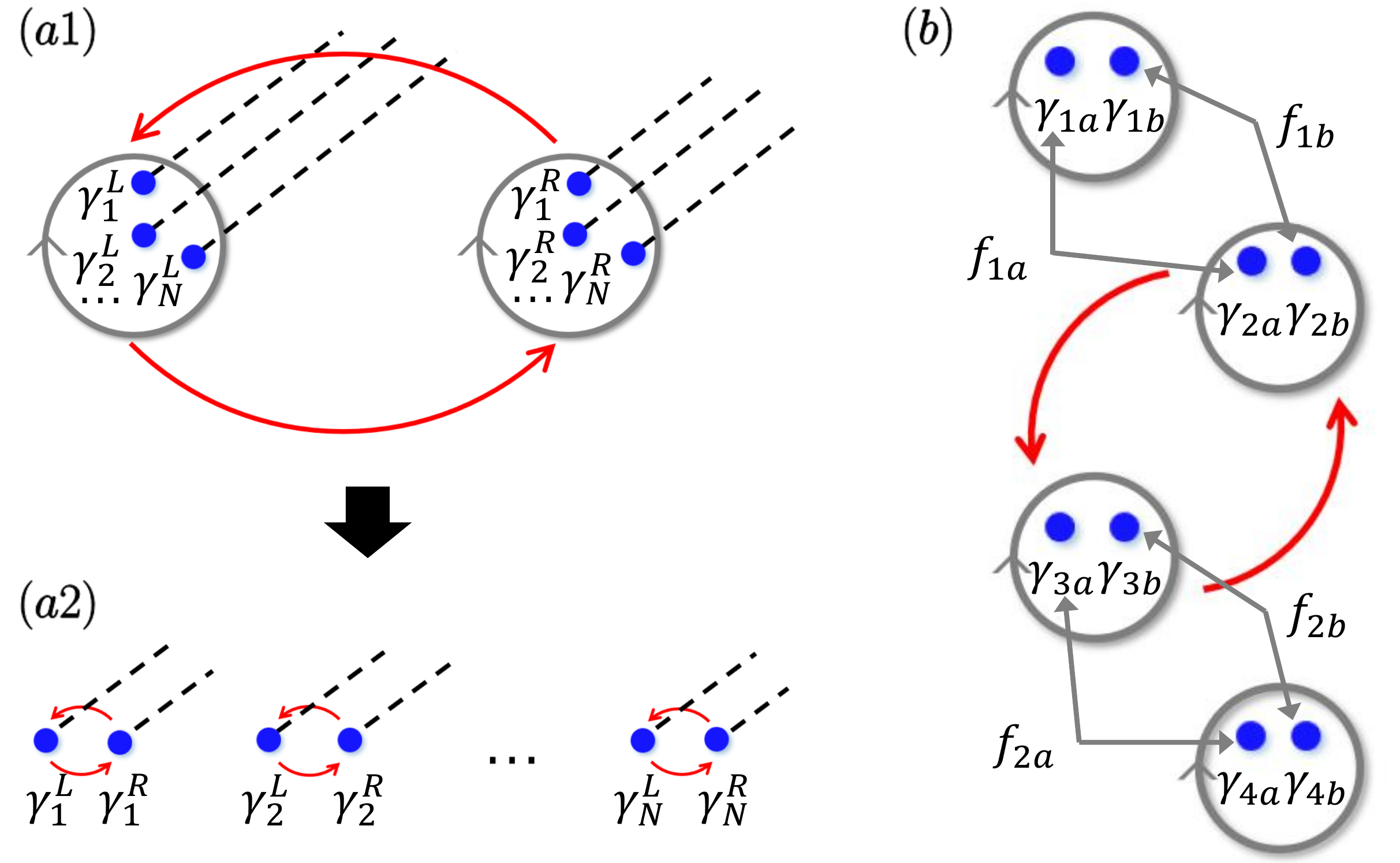}
\par\end{centering}
\caption{Schematic diagrams for MZMs residing at vortices in a unitary symmetry-protected TSC. The gray circles with arrows denote vortices, the blue dots are MZMs, and the dashed lines represent the branch cuts of MZMs. (a) The braiding of multiple pairs of MZMs is reduced into mulitple individual sectors, each of which two MZMs independently. (b) The sketch of $\mathcal{B}_{23}$ in the concrete model. The local MZMs $\gamma_{ia}$ and $\gamma_{ib}$ define a complex fermion $\eta_{i}$. The non-local complex ferimons $f_{ia(b)}$ are defined by MZMs in different vortices. }\label{sketch}
\end{figure}

\textit{Generic theory.--}We start with the generic theory of the non-Abelian statistics for unitary symmetry-protected MZMs. Consider multiple MZMs localized at each edge or vortex core in a symmetry-protected TSC, and the couplings among them are excluded by unitary symmetries~\cite{Kotetes2013,GilbertBernevig2014,XJLiu2014C4,
Ryu2016review,Nagaosa2014}. The braiding of two vortices exchanges the multiple pairs of MZMs as shown schematically in Fig.~\ref{sketch}(a1).
Let the system described by the Hamiltonian $H$ have a unitary symmetry $\mathcal{R}$, i.e., $\mathcal{R}H\mathcal{R}^{-1}=H$ and $\mathcal{R}i\mathcal{R}^{-1}=i$, with $N$ symmetry-protected MZMs at both the left and right hand side edges
or vortices. The symmetry protection implies that $\mathcal{R}i\tilde{\gamma}_{i}^{p}\tilde{\gamma}_{j}^{p}\mathcal{R}^{-1}=-i\tilde{\gamma}_{i}^{p}\tilde{\gamma}_{j}^{p}$ so that the coupling between two MZMs $\tilde{\gamma}_{i,j}$ in the same vortex core (left for $p=L$ or right for $p=R$) is forbidden. The braiding is a unitary dynamical evolution with duration $T$ which can be described by time-ordering integral $U(T)=\hat{T}\exp\left[{-i\int_{-T/2}^{T/2}\mathrm{d}\tau H(\tau)}\right]$.
When the TSC satisfies unitary symmetry at all time $\mathcal{R}H(t)\mathcal{R}^{-1}=H(t)$ for $-T/2<t<T/2$, the unitary evolution also satisfies the symmetry $[\mathcal{R},U(T)]=0$ (see more details in Supplementary Material~\cite{SM}).
We define the effective braiding Hamiltonian $H_{E}$ via $H_{E}=iT^{-1}\ln U(T)$ to characterize the braiding. 
Since the operator $U(T)$ represents a unitary transformation of MZMs, the braiding
Hamiltonian $H_{E}$ is linear and takes the generic form $H_{E}=i\sum_{ij}\epsilon_{ij}\tilde{\gamma}_{i}^{L}\tilde{\gamma}_{j}^{R}$ due to  unitary symmetry, which can be written in an off-diagonal matrix form in the bases $\{|\gamma_{1}^{L}\rangle,\cdots,|\gamma_{N}^{L}\rangle;|\gamma_{1}^{R}\rangle,\cdots|\gamma_{N}^{R}\rangle\}$. The mode $|\gamma_j^p\rangle$ is a proper linear transformation of $|\tilde{\gamma}^p_i\rangle$~\cite{SM}. The unitary braiding operator can then be proven to be $U = \bigl(\begin{smallmatrix}0 & -1\\ 1 & 0\end{smallmatrix}\bigr)$, where $0$ and $1$ are $N\times N$ zero and identity matrices, respectively~\cite{SM}. It follows that $U\gamma_{i}^{L(R)}=\mathcal{B}\gamma_{i}^{L(R)}\mathcal{B}^{-1}$, where the braiding operator
\begin{equation}
\mathcal{B}=\mathrm{e}^{-\frac{\pi}{4}\gamma_{1}^{L}\gamma_{1}^{R}}\mathrm{e}^{-\frac{\pi}{4}\gamma_{2}^{L}\gamma_{2}^{R}} \cdots\mathrm{e}^{-\frac{\pi}{4}\gamma_{N}^{L}\gamma_{N}^{R}}.
\end{equation}
This result shows that braiding two vortices generically reduces to $N$ independent sectors, with each sector braiding two different Majorana modes due to the symmetry protection [Fig.~\ref{sketch}(a2)]. In other words, during the braiding {\em a MZM in one vortex can only effectively see one, rather than all MZMs in another vortex.}

\begin{figure}
\begin{centering}
\includegraphics[width=\columnwidth]{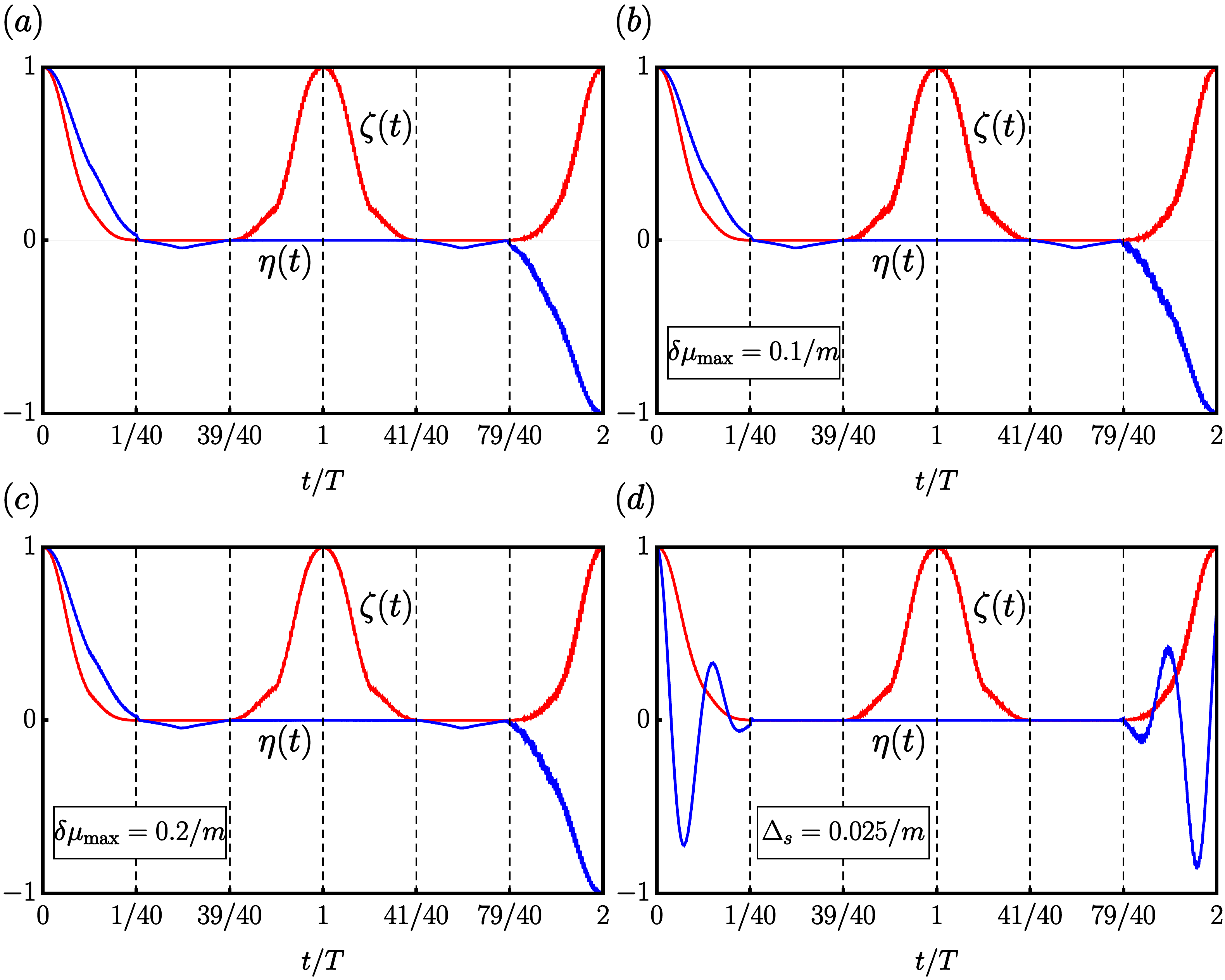}
\par\end{centering}
\caption{Numerical results of braiding two vortices with perturbations. (a) Evolution of MZM wave function in a full braiding. (b-c) Evolution with symmetry-preserving randomized chemical potential $\delta\mu$. (d) Evolution with symmetry-breaking s-wave pairing $\Delta_s$. The non-Abelian braiding is confirmed in (a-c) that $\eta(t)=\langle\gamma_{1a}|\gamma_{1a}(t)\rangle|_{t=2T}=-1$ after a full braiding at $t=2T$ (similar for other MZMs), and breaks down in (d). The adiabatic condition is satisfied in that $\zeta(t)=\sum_{j=1,2}[|\langle\gamma_{1a}(t)|\gamma_{ja}(0)\rangle|^2+\langle\gamma_{1a}(t)|\gamma_{jb}(0)\rangle|^2]$ returns to unity after a single ($t=T$) and a full ($t=2T$) braiding.
Here $\mu=1/m$ and $\Delta=1/2m$.}\label{numgamma}
\end{figure}

\begin{table*}[htb]
	\renewcommand\arraystretch{1.5}
	\centering
      \caption{Examples for symmetry-preserving or symmetry-breaking terms in TSC and QAHI. The letter ``P'' (``B'') in the rightmost column means that the term preserves (breaks) the symmetry that protects the topology.}
\label{symmetry}
\hspace{0cm}
\begin{tabular}{p{6cm}<{\centering}p{6cm}<{\centering}p{0.8cm}<{\centering}p{0.8cm}<{\centering}p{0.8cm}<{\centering}p{1.3cm}<{\centering}}
    \hline\hline
    TSC & QAHI & $\mathcal{S}_1$ & $\mathcal{S}_2$ & $\mathcal{S}_1\cdot\mathcal{S}_2$ & Topology\\
    \hline
    $\sum_{\bm{i}}W_{\bm{i}}(n_{\bm{i}\uparrow}+n_{\bm{i}\downarrow})$ & $\sum_{\bm{i}}W_{\bm{i}}(n_{\bm{i}\uparrow}-n_{\bm{i}\downarrow})$  &   \cmark & \cmark &  \cmark & P \\
    $\sum_{j_{x}}t_{\text{SO}}(c_{j_{x}\uparrow}^{\dagger}c_{j_{x}+1\downarrow}-c_{j_{x}\uparrow}^{\dagger}c_{j_{x}-1\downarrow})+\text{h.c.}$ &  $\sum_{j_{x}}\Delta_{p}(c_{j_{x}\uparrow}^{\dagger}c_{j_{x}+1\downarrow}^{\dagger}-c_{j_{x}\uparrow}^{\dagger}c_{j_{x}-1\downarrow}^{\dagger})+\text{h.c.}$ & \xmark &  \cmark &  \xmark & P \\
    $i\sum_{j_{y}}t_{\text{SO}}(c_{j_{y}\uparrow}^{\dagger}c_{j_{y}+1\downarrow}-c_{j_{y}\uparrow}^{\dagger}c_{j_{y}-1\downarrow})+\text{h.c.}$ &  $i\sum_{j_{y}}\Delta_{p}(c_{j_{y}\uparrow}^{\dagger}c_{j_{y}+1\downarrow}^{\dagger}-c_{j_{y}\uparrow}^{\dagger}c_{j_{y}-1\downarrow}^{\dagger})+\text{h.c.}$ &  \cmark & \xmark  & \xmark & P \\
    $\sum_{\bm{i}}B_{x}(c_{\bm{i}\uparrow}^{\dagger}c_{\bm{i}\downarrow}+c_{\bm{i}\downarrow}^{\dagger}c_{\bm{i}\uparrow})$ &  $\sum_{\bm {i}}\Delta_{s}(c_{\bm{i}\uparrow}^{\dagger}c_{\bm{i}\downarrow}^{\dagger}+\text{h.c.})$ &  \cmark & \xmark  & \xmark & P \\
    $\sum_{\bm{i}}iB_{y}(-c_{\bm{i}\uparrow}^{\dagger}c_{\bm{i}\downarrow}+c_{\bm{i}\downarrow}^{\dagger}c_{\bm{i}\uparrow})$ &  $\sum_{\bm{i}}\Delta_{s}(ic_{\bm{i}\uparrow}^{\dagger}c_{\bm{i}\downarrow}^{\dagger}+\text{h.c.})$ & \xmark &  \cmark &  \xmark  & P \\
    $\sum_{\bm{ij},\sigma=\uparrow,\downarrow}\Delta_{p}c_{\bm{i}\sigma}^{\dagger}c_{\bm{j}\sigma}^{\dagger}+\text{h.c.}$ &  $\sum_{\bm{ij}}\Delta_{p}(c_{\bm{i}\uparrow}^{\dagger}c_{\bm{j}\uparrow}^{\dagger}-c_{\bm{i}\downarrow}^{\dagger}c_{\bm{j}\downarrow}^{\dagger})+\text{h.c.}$ &  \cmark &  \xmark & \xmark & P \\
    $\sum_{\bm{ij}}\Delta_{p}(c_{\bm{i}\uparrow}^{\dagger}c_{\bm{j}\uparrow}^{\dagger}-c_{\bm{i}\downarrow}^{\dagger}c_{\bm{j}\downarrow}^{\dagger})+\text{h.c.}$ &  $\sum_{\bm{ij},\sigma=\uparrow,\downarrow}\Delta_{p}c_{\bm{i}\sigma}^{\dagger}c_{\bm{j}\sigma}^{\dagger}+\text{h.c.}$ &  \xmark & \cmark & \xmark & P \\
    $\sum_{\bm{i}}\Delta_{s}c_{\bm{i}\uparrow}^{\dagger}c_{\bm{i}\downarrow}^{\dagger}+\text{h.c.}$ &  $\sum_{\bm{i}}Bc_{\bm{i}\uparrow}^{\dagger}c_{\bm{i}\downarrow}+\text{h.c.}$ & \xmark &  \xmark & \cmark & B \\
    \hline\hline
  \end{tabular}
\end{table*}

\textit{TSC Model.--}We apply the generic theory to a $p_{x}+ip_{y}$ TSC, with the Cooper pairs in the state $S=1$ and $S_z=0$. The Hamiltonian reads $H_{\text{BdG}} = \frac{1}{2} \sum_{\bm{p}} \Psi_{\bm{p}}^\dagger  \mathcal{H}_{\text{BdG}} \Psi_{\bm{p}}$, where $\Psi_{\bm{p}}=(c_{\bm{p}\uparrow},c_{\bm{p}\downarrow},c_{-\bm{p}\uparrow}^{\dagger},c_{-\bm{p}\downarrow}^{\dagger})^{\text{T}}$, and
\begin{eqnarray}
\mathcal{H}_{\text{BdG}}&=&(p^2/2m-\mu)\tau_z\otimes\sigma_0-2|\Delta|(p_x\tau_y+p_y\tau_x)\otimes\sigma_x \nonumber\\
&\equiv& h_z \tau_z\otimes\sigma_0 + h_y \tau_y\otimes\sigma_x + h_x \tau_x\otimes\sigma_x.\label{Ham}
\end{eqnarray}
Here $\tau_{i}$ and $\sigma_{i}$ act on particle-hole and spin space, respectively. 
The bulk spectra $E(\bm{p})=\pm|\bm{h}|$ are gapless if $\mu=0$. This model possesses unitary symmetries $\mathcal{S}_1=\tau_0\otimes\sigma_x$, $\mathcal{S}_2=\tau_z\otimes\sigma_y$ and $\mathcal{S}_1 \mathcal{S}_2=\tau_{z}\otimes i\sigma_{z}$. The MZMs in vortex are protected by $\mathcal{S}_1$ and $\mathcal{S}_2$, but not by $\mathcal{S}_3$. The first seven rows in Table~\ref{symmetry} show various perturbations that keep $\mathcal{S}_1$ or $\mathcal{S}_2$. The last row shows a perturbation that breaks $\mathcal{S}_{1,2}$ while not $\mathcal{S}_1\mathcal{S}_2$, but the MZMs in vortex are not protected. The bulk Chern number computed via $\hat{\bm{h}}=\bm{h}/|\bm{h}|$ is
$C=\int\frac{\mathrm{d}\bm{p}}{2\pi}[\hat{\bm{h}}\cdot(\partial_{p_{x}}\hat{\bm{h}}\times\partial_{p_{y}}\hat{\bm{h}})]=-2$ for $\mu>0$.

Two MZMs exist in each vortex for $C=-2$ phase and obey non-Abelian statistics. The vortex can be introduced by taking $|\Delta|\rightarrow|\Delta|\cdot g(r)\mathrm{e}^{-i\theta}$, where $g(r)$ depends on $r$ and $(r,\theta)$ are the polar coordinates. The vortex phase can be gauge-transformed into the operators via $c_{\uparrow(\downarrow)}(r,\theta)\to c_{\uparrow(\downarrow)}(r,\theta)\mathrm{e}^{-\mathrm{i}\theta/2}$, which then obey the anti-periodic boundary condition. The two MZMs read
\begin{eqnarray}
\gamma_{a}&= & \frac{1}{\mathcal{N}}\int r\mathrm{d}r\mathrm{d}\theta f(r)[-\mathrm{e}^{i\theta/2}c_{\uparrow}(r,\theta)+\mathrm{e}^{i\theta/2}c_{\downarrow}(r,\theta)\nonumber\\
 && -\mathrm{e}^{-i\theta/2}c_{\uparrow}^{\dagger}(r,\theta)+\mathrm{e}^{-i\theta/2}c_{\downarrow}^{\dagger}(r,\theta)],
\end{eqnarray}
and $\gamma_{b}=\mathcal{S}_2\gamma_{a}\mathcal{S}_2^\dagger$, where $\mathcal{N}$ is the normalization factor, 
and $f(r)=\frac{1}{\sqrt{r}}\exp\left[{-\frac{1}{2}\int_{0}^{r}\frac{\mu}{|\Delta|g(r')}\mathrm{d}r'}\right]$. The two MZMs are protected by the unitary symmetry $\mathcal{S}_{1}$ or $\mathcal{S}_{2}$, which also protects the non-Abelian statistics. The non-Abelian braiding and symmetry protection are confirmed by numerical simulation shown in Fig.~\ref{numgamma} through a lattice Hamiltonian (see Supplemental Material~\cite{SM}). As predicted, each MZM (e.g. $\gamma_{1a}$) in braiding accumulates a $\pi$ phase after a full braiding [Fig.~\ref{numgamma}(a)]. This braiding result is robust against randomized chemical potential $\mu'({\bm r})$ with $|\mu'({\bm r})-\mu|\leq\delta\mu_{\text{max}}$, which preserves the symmetries [Fig.~\ref{numgamma}(b-c)]. Similar results are expected for other disorders without breaking the $\mathcal{S}_{1}$ or $\mathcal{S}_{2}$ symmetry. In comparison, a symmetry-breaking $s$-wave superconducting order vitiates the braiding.

There are different ways to introduce the complex fermion modes from the MZMs, as illustrated in Fig.~\ref{sketch} (b). One is to define the complex fermion modes in non-local bases $f_{ia(b)}=(\gamma_{2i-1a(b)}+i\gamma_{2ia(b)})/2$, and another is from local MZMs in each vortex as $\eta_{i}=(\gamma_{ia}+i\gamma_{ib})/2$. (i) In the $f$-representation, the MZMs are classified into two sectors denoted by $a$ and $b$, and
the braiding operators can be generally decomposed into a direct product of the matrices of each sector.
For example, for four vortices, the Fock space is spanned by $|n_{1a}n_{1b}\rangle_{f}|n_{2a}n_{2b}\rangle_{f}$, which denotes states with particle number $n_{ia(b)}$ for $f_{ia(b)}$.
In the subspace \{$|00\rangle_{f}|00\rangle_{f}$,$|01\rangle_{f}|01\rangle_{f}$,$|10\rangle_{f}|10\rangle_{f}$,$|11\rangle_{f}|11\rangle_{f}$\},
the braiding matrix of $\mathcal{B}_{23}$ is
\begin{equation}
B_{23}^{(f)}=\frac{1}{\sqrt{2}}\begin{pmatrix}
1 & -i\\
-i & 1
\end{pmatrix}\otimes\frac{1}{\sqrt{2}}\begin{pmatrix}
1 & -i\\
-i & 1
\end{pmatrix},
\end{equation}
which connects states with the same fermion parity in each sector. (ii) In the $\eta$-representation, there is a ``particle-number conservation'' due to the relation $\mathcal{B}_{ij}=1+\eta_{j}^{\dagger}\eta_{i}-\eta_{i}^{\dagger}\eta_{j}-n_{i}-n_{j}+2n_{i}n_{j}$, with $n_{i}=\eta_{i}^{\dagger}\eta_{i}$. We initialize the non-local fermion states as follows.  Consider $2N$ vortices (or anti-vortices) in TSC and couple them via $H_{2i-1,2i}=\lambda\mathrm{e}^{i\psi_{2i-1,2i}}\eta_{2i-1}^\dagger\eta_{2i}+\text{h.c.}$. Here the phases of $\eta_{2i-1}$ and $\eta_{2i}$ have been absorbed into $\psi_{2i-1,2i}$, and the eigenstates are $\eta_{i\pm}=  \frac{1}{\sqrt{2}}\left(\eta_{2i-1}\pm\mathrm{e}^{-i\psi_{2i-1,2i}}\eta_{2i}\right)$. The braiding evolves the state of the system in the particle-number conserved subspaces. For example, in
the single-particle subspace for four vortices \{$|10\rangle_{\eta}|00\rangle_{\eta}$,$|01\rangle_{\eta}|00\rangle_{\eta}$,$|00\rangle_{\eta}|10\rangle_{\eta}$,$|00\rangle_{\eta}|01\rangle_{\eta}$\},
where  $|n_{1-}n_{1+}\rangle_{\eta}|n_{2-}n_{2+}\rangle_{\eta}$ denotes states with particle number $n_{i\pm}$ for $\eta_{i\pm}$,
the matrix of operator $\mathcal{B}_{23}$ is
\begin{equation}
B_{23}^{(\eta)}=\frac{1}{2}\begin{pmatrix}
1 & 1 & \mathrm{e}^{-i\psi_{12}} & \mathrm{e}^{-i\psi_{12}}\\
1 & 1 & -\mathrm{e}^{-i\psi_{12}} & -\mathrm{e}^{-i\psi_{12}}\\
-\mathrm{e}^{i\psi_{12}} & \mathrm{e}^{i\psi_{12}} & 1 & -1\\
-\mathrm{e}^{i\psi_{12}} & \mathrm{e}^{i\psi_{12}} & -1 & 1
\end{pmatrix}.\label{B23eta}
\end{equation}
Under $B_{23}^{(\eta)}$, each basis vector evolves into the superposition
of all the single-particle states. Note that $f$ is generally a linear
combination of $\eta$ and $\eta^{\dagger}$, and the particle-number conservation for $\eta$ and the parity conservation for $f$ are not contradictive.

\begin{figure}
\begin{centering}[t]
\includegraphics[width=1\columnwidth]{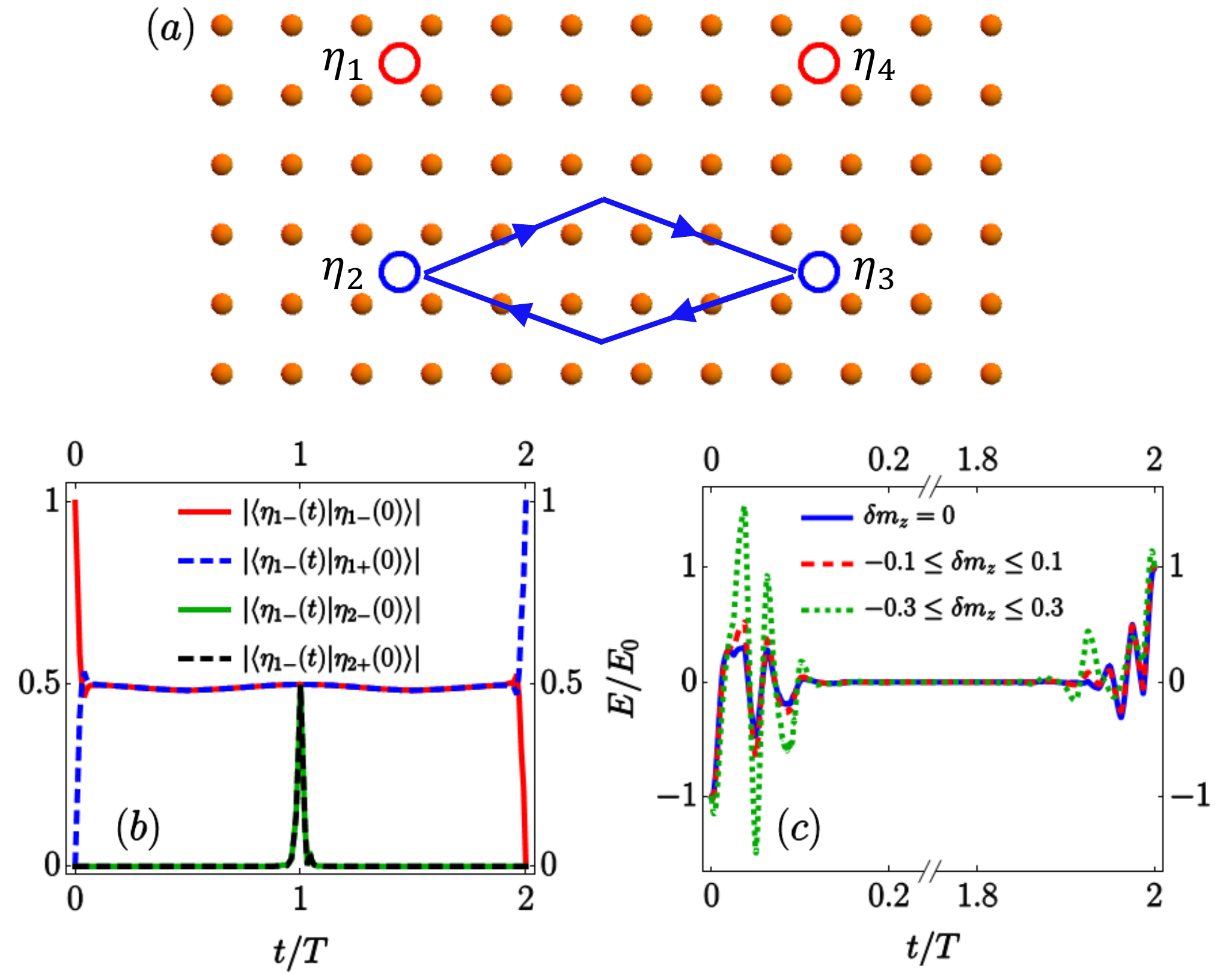}
\par\end{centering}
\caption{Numerical simulation of a full braiding between $\eta_{2}$ and $\eta_{3}$. (a) Orange ball denotes the lattice
sites, and Vortices (anti-vortices) are denoted by blue (red) circles. (b) The evolution of wave function $|\eta_{1-}(t)\rangle$. The system evolves into a superposition of the four single particle states as braiding
$\eta_{2}$ and $\eta_{3}$ once. After a full braiding, $|\eta_{1-}(t)\rangle$
evolves into $|\eta_{1+}\rangle$. (c) The energy $E$ of $|\eta_{1-}(t)\rangle$ as a function of time. The absolute
value of $|\eta_{1-}(0)\rangle$ is denoted by $E_{0}$. The
dashed lines show the energy evolution with certain randomization
$\delta m_{z}$ in the $m_{z}$ term. After a full braiding, all the
three curves arrive at the positive energy $E_{0}$, demonstrating
the robustness of non-Abelian statistics.
Here we set $m_{z}=2t_0$ and $t_{\text{SO}}=t_0$.}\label{numbraiding}
\end{figure}

\textit{Simulation of the non-Abelian statistics by QAHI.--} We show now a highly nontrivial result that the unitary symmetry-protected non-Abelian statistics of MZM pairs can be mapped to and simulated by vortex modes in a QAHI. After the transformation $c_{\downarrow}\leftrightarrow c_{\downarrow}^{\dagger}$ and the replacements $|\Delta|\cdot g(r)\mathrm{e}^{-i\theta}\to t_{\text{SO}}\cdot g(r)\mathrm{e}^{-i\theta}$ and $\mu\to-m_{z}$, the Hamiltonian (\ref{Ham}) describes a QAHI with vortices:
\begin{eqnarray}
\mathcal{H}_{\text{QAHI}}(r,\theta)&=&\left[t_0\left(r^{-1}\partial_{r}r\partial_r+r^{-2}\partial_{\theta}^2\right)+m_z\right]\sigma_z\nonumber\\
&&+2it_{\rm SO}g(r)\mathrm{e}^{-i\theta}(r^{-1}\sigma_r\partial_\theta+\sigma_\theta\partial_r),\label{QAHI}
\end{eqnarray}
where $\sigma_r=\cos\theta\sigma_x+\sin\theta\sigma_y$ and $\sigma_\theta=\cos\theta\sigma_y-\sin\theta\sigma_x$. The operator that annihilates the zero mode is just $\eta=(\gamma_a+i\gamma_b)/2$ after particle-hole transformation. Noted that our realization of zero modes in a QAHI is different from the traditional proposal by digging a hole with a $\pi$ flux threading through~\cite{ShenBook}. In above Hamiltonian~(\ref{QAHI}), we only introduce a phase vortex for $t_{\rm SO}$. Similar to the case of the TSC model, the topology is protected by $\mathcal{S}_1=\tau_x\otimes\sigma_x$ and $\mathcal{S}_2=\tau_y\otimes\sigma_x$. Table~\ref{symmetry} shows examples of perturbation that keep or break the symmetry. To illustrate the nontrivial braiding statistics, we consider four vortices (or anti-vortices) in a QAHI so that there are four zero modes $\eta_{1}$, $\eta_{2}$, $\eta_{3}$ and $\eta_{4}$ [see Fig.~\ref{numbraiding}(a)]. We initialized the system in the state $|\eta_{1-}\rangle$ and perform a numerical simulation in position space, as shown in Fig.~\ref{numbraiding}(b-c). In the braiding, $\eta_2$ and $\eta_3$ travels away from their initial locations such that $|\eta_{1-}(t)\rangle$ only overlaps with $|\eta_{1-}(0)\rangle$ or $|\eta_{1+}(0)\rangle$ [Fig.~\ref{numbraiding}(b)]. After braiding once, the state $|\eta_{1-}(t=T)\rangle$ evolves into a linear combination of the four states as expected in Eq.~(\ref{B23eta}), and after a full braiding, the state $|\eta_{1-}(t=2T)\rangle$ evolves into $|\eta_{1+}(0)\rangle$. Further, we demonstrate the robustness of non-Abelian braiding against perturbations which keep the symmetries. As shown in Fig.~\ref{numbraiding}(c), the full braiding transforms the negative energy state $|\eta_{1-}(t)\rangle$ into the positive energy one $|\eta_{1+}(t)\rangle$; the disorder in $m_z$ only induces fluctuations but cannot affect the braiding result.

We can realize various topological quantum gates by braiding the vortices. Due to the particle-number conservation, the realization of quantum logic gates in a QAHI is restricted in the $\eta$-representation. For different subspaces, there are different methods to realize quantum logic gates. For example, in the two-particle subspace and defining $|00\rangle=|11\rangle_{\eta}|00\rangle_{\eta}$, $|10\rangle=|00\rangle_{\eta}|11\rangle_{\eta}$, $|01\rangle=|10\rangle_{\eta}|01\rangle_{\eta}$, $|11\rangle=|01\rangle_{\eta}|10\rangle_{\eta}$, we have
\begin{equation}
\text{Z-gate: }B_{12}^{(\eta)\,2}=B_{34}^{(\eta)\,2},\quad\text{CNOT-gate: }-B_{23}^{(\eta)\,2}.
\end{equation}
More vortices can be added to provide ancilla qubits. When considering six
vortices, one can define $|00\rangle=|10\rangle_{\eta}|00\rangle_{\eta}|00\rangle_{\eta}$, $|10\rangle=|01\rangle_{\eta}|00\rangle_{\eta}|00\rangle_{\eta}$, $|01\rangle=|00\rangle_{\eta}|10\rangle_{\eta}|00\rangle_{\eta}$, $|11\rangle=|00\rangle_{\eta}|01\rangle_{\eta}|00\rangle_{\eta}$ as the four practical
qubits (the remaining two single-particle states serve as ancilla ones). We then have
\begin{align}
\begin{split}
\text{Z-gate: }&B_{34}^{(\eta)\,2},\hspace*{40pt}\text{CNOT-gate: }B_{45}^{(\eta)\,2}, \\
\text{X-gate: }&B_{23}^{(\eta)}B_{34}^{(\eta)}B_{12}^{(\eta)}B_{23}^{(\eta)},\;\text{when }\psi_{12}=\psi_{34}.
\end{split}
\end{align}
Compared with previous realization of quantum logic gates utilizing
Majorana qubits~\cite{Pachos2017}, the current proposal provides simpler methods to realize CNOT-gate.

\begin{figure}
\begin{centering}
\includegraphics[width=\columnwidth]{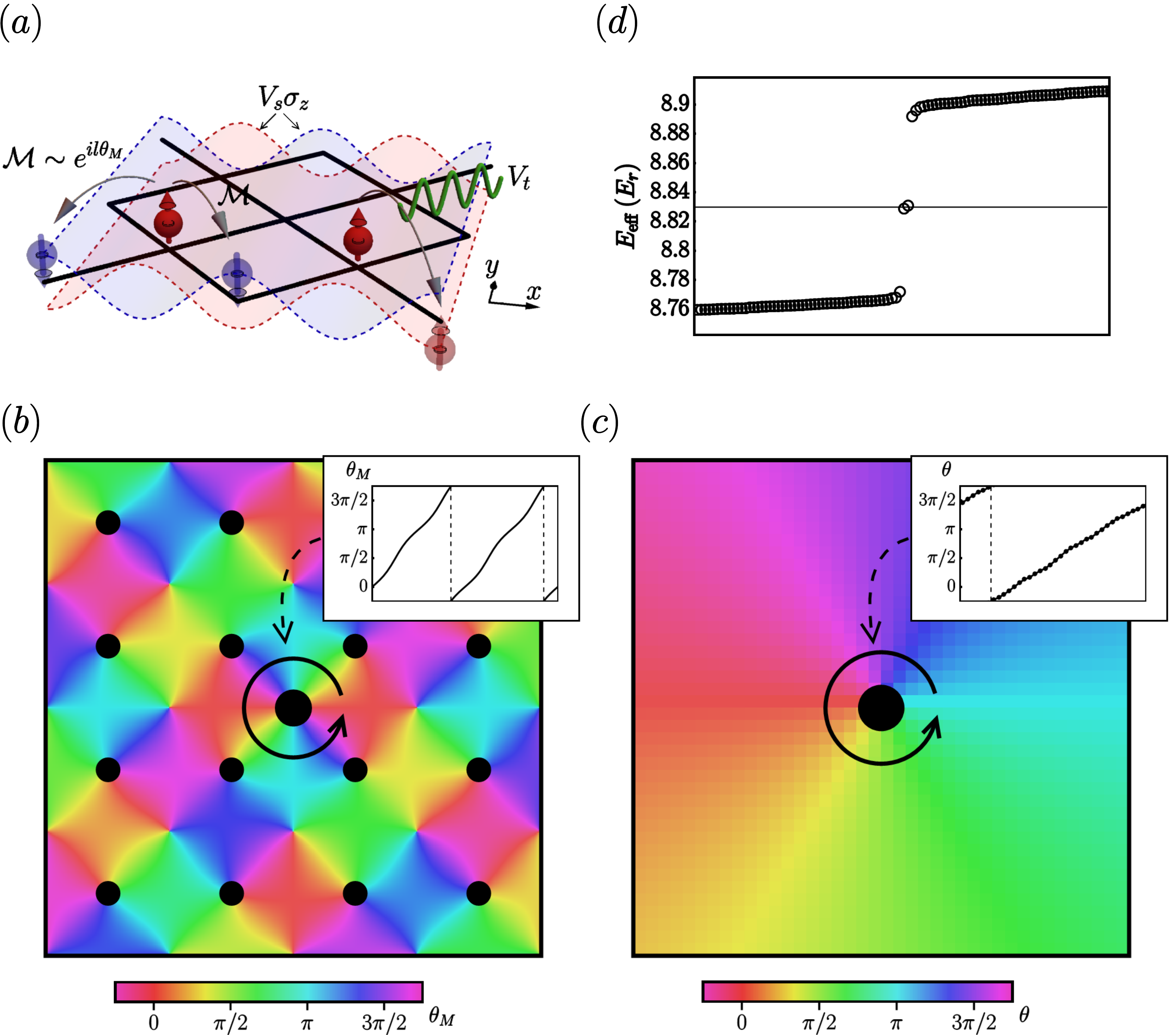}
\par\end{centering}
\caption{(a) Sketch of the experimental realization scheme. The atoms reside at a primary checkerboard lattice (black lines) with  a staggered onsite Zeeman potential $V_s$.
The Raman potential ${\cal M}$ induces the spin-flip nearest neighbor hopping. A shaking lattice $V_t$ is applied to drive the spin-conserved hopping by compensating neighboring onsite energy difference. (b) The phase of Raman potential $\theta_M$ with $l=1$. (c) The phase of the spin-flip hopping $\theta$ in the $x$-direction produced by the Raman potential with $l=1$. (d) The energy spectrum of the system with a vortex ($l=1$) and an anti-vortex ($l=-1$). See details in ref.~\cite{SM}.}\label{QAHSim}
\end{figure}

\textit{Experimental realization.}--Finally we propose a feasible experimental setup to simulate a QAHI with vortices via an optical Raman lattice~\cite{ORL2014,ORL2016,ORL2019,ORL2020}. The realization scheme is depicted in Fig.~\ref{QAHSim}(a), and the details can found in the Supplemental Material~\cite{SM}. Besides the primary square lattice, there are three main ingredients in our proposal:
(i) Raman potential $\mathcal{M}$ with an orbital angular momentum $l$, which induces spin-flip hopping $t_{\rm SO}$-term and the phase vortex. The orbital angular momentum can be imprinted by applying an Laguerre-Gaussian (LG) beam~\cite{LG1,LG2}.
(ii) Staggered on-site Zeeman term $V_s \sigma_z$, which provides an on-site energy difference between the two spin states so that on-site spin flipping is suppressed.
(iii) Shaking lattice $V_t$, which is applied to induce spin-conserved nearest-neighbor hopping. 

We confirm by exact numerical calculations that such a realization indeed simulates a QAHI with vortices described by the Hamiltonian (\ref{QAHI}). We first calculate the phase of Raman potential $\theta_M$ in the case $l=1$.
As in Fig.~\ref{QAHSim}(b), the winding of the phase $\theta_M$ is $2$ along a path encircling the center of LG beam, while for any other paths encircling one center of lattice barrier, the winding is $1$. The additional winding causes the spin-flip hopping $t_{\rm SO}$ to have a phase vortex at the center of the LG beam, as numerically demonstrated in Fig.~\ref{QAHSim}(c).
We then calculate the energy spectrum of the system with a vortex ($l=1$) and an anti-vortex ($l=-1$)~\cite{SM}.
The result is shown in Fig.~\ref{QAHSim}(d). One can find two nearly degenerate states at the center of the gap, which confirms the existence of zero modes in vortices. The braiding of zero modes can then be realized by slowly moving the LG beam. Details of the realization are seen in ref.~\cite{SM}.

\textit{Conclusion.}--We have uncovered that Majorana modes protected by unitary symmetries generically obey symmetry protected non-Abelian statistics, and proposed the experimental realization. 
Specifically, the MZM pairs in vortex cores of a unitary symmetry-protected spin-triplet TSC exemplify the non-Abelian statistics with analytical and numerical results. In comparison with TSCs without symmetry-protection, the Majorana qubit states with unitary symmetry-protection can be defined in the local and non-local bases, which facilitate the realization of various topological quantum gates. We further mapped the spin-triplet TSC to QAHI, with which we proposed a practical experimental scheme in optical Raman lattice to verify the present study. This work enriches the braiding statistics and may open up diverse routes toward realizing topological quantum computers.


\begin{thebibliography}{99}


\bibitem{Kitaev2001} A. Y. Kitaev, Physics-Uspekhi {\bf 44}, 131 (2001).

\bibitem{ReadGreen2000} N. Read and D. Green, Phys. Rev. B {\bf 61}, 10267 (2000).

\bibitem{Nayak2008} C. Nayak, S. H. Simon, A. Stern, M. Freedman, and S. Das Sarma, Rev. Mod. Phys. {\bf 80}, 1083 (2008).

\bibitem{Nayak1996} C. Nayak and F. Wilczek, Nucl. Phys. B {\bf 3}, 529 (1996).

\bibitem{Ivanov2001} D. A. Ivanov, Phys. Rev. Lett. {\bf 86}, 268 (2001).

\bibitem{DasSarma2005} S. Das Sarma, M. Freedman, and C. Nayak, Phys. Rev. Lett. {\bf 94}, 166802 (2005).

\bibitem{Alicea2011} J. Alicea, Y. Oreg, G. Refael, F. Von Oppen, and M. P. Fisher, Nat. Phys. {\bf 7}, 412 (2011).

\bibitem{Kitaev2003} A. Kitaev, Annals of Physics {\bf 303}, 2 (2003).

\bibitem{Aasen2016} D. Aasen, M. Hell, R. V. Mishmash, A. Higginbotham, J. Danon, M. Leijnse, T. S. Jespersen, J. A. Folk, C. M. Marcus, K. Flensberg, and J. Alicea, Phys. Rev. X {\bf 6}, 031016 (2016).

\bibitem{Pachos2017} V. Lahtinen and J. K. Pachos, SciPost Phys. {\bf 3}, 021 (2017).

\bibitem{Mourik2012} V. Mourik, K. Zuo, S. M. Frolov, S. Plissard, E. P. Bakkers, and L. P. Kouwenhoven, Science {\bf 336}, 1003 (2012).

\bibitem{MTDeng2012} M. Deng, C. Yu, G. Huang, M. Larsson, P. Caroff, and H. Xu, Nano Letters {\bf 12}, 6414 (2012).

\bibitem{Rokhinson2012} L. P. Rokhinson, X. Liu, and J. K. Furdyna, Nat. Phys. {\bf 8}, 795 (2012).

\bibitem{Shtrikman2012} A. Das, Y. Ronen, Y. Most, Y. Oreg, M. Heiblum, and H. Shtrikman, Nat. Phys. {\bf 8}, 887 (2012).

\bibitem{JFJia2012} M.-X. Wang, C. Liu, J.-P. Xu, F. Yang, L. Miao, M.-Y. Yao, C. L. Gao, C. Shen, X. Ma, X. Chen, Z.-A. Xu, Y. Liu, S.-C. Zhang, D. Qian, J.-F. Jia, and Q.-K. Xue, Science {\bf 336}, 52 (2012).

\bibitem{Marcus2013} H. O. H. Churchill, V. Fatemi, K. Grove-Rasmussen, M. T. Deng, P. Caroff, H. Q. Xu, and C. M. Marcus, Phys. Rev. B {\bf 87}, 241401 (2013).

\bibitem{JFJia2014} J.-P. Xu, C. Liu, M.-X. Wang, J. Ge, Z.-L. Liu, X. Yang, Y. Chen, Y. Liu, Z.-A. Xu, C.-L. Gao, D. Qian, F.-C. Zhang, and J.-F. Jia, Phys. Rev. Lett. {\bf 112}, 217001 (2014).

\bibitem{Yazdani2014} S. Nadj-Perge, I. K. Drozdov, J. Li, H. Chen, S. Jeon, J. Seo, A. H. MacDonald, B. A. Bernevig, and A. Yazdani, Science {\bf 346}, 602 (2014).

\bibitem{Marcus2015} W. Chang, S. Albrecht, T. Jespersen, F. Kuemmeth, P. Krogstrup, J. Nyg{\aa}rd, and C. M. Marcus, Nature Nanotechnology {\bf 10}, 232 (2015).

\bibitem{Marcus2016} S. M. Albrecht, A. P. Higginbotham, M. Madsen, F. Kuemmeth, T. S. Jespersen, J. Nyg{\aa}rd, P. Krogstrup, and C. Marcus, Nature {\bf 531}, 206 (2016).

\bibitem{Molenkamp2016} J. Wiedenmann, E. Bocquillon, R. S. Deacon, S. Hartinger, O. Herrmann, T. M. Klapwijk, L. Maier, C. Ames, C. Br{\"u}ne, C. Gould, A. Oiwa, K. Ishibashi, S. Tarucha, H. Buhmann, and L. W. Molenkamp, Nature Communications {\bf 7}, 1 (2016).

\bibitem{Molenkamp2017} E. Bocquillon, R. S. Deacon, J. Wiedenmann, P. Leubner, T. M. Klapwijk, C. Br{\"u}ne, K. Ishibashi, H. Buhmann, and L. W. Molenkamp, Nature Nanotechnology {\bf 12}, 137 (2017).

\bibitem{Qiao2017} K. Zhang, J. Zeng, Y. Ren, and Z. Qiao, Phys. Rev. B {\bf 96}, 085117 (2017).

\bibitem{HDing2018a} P. Zhang, K. Yaji, T. Hashimoto, Y. Ota, T. Kondo, K. Okazaki, Z. Wang, J. Wen, G. D. Gu, H. Ding, and S. Shin, Science {\bf 360}, 182 (2018).

\bibitem{HDing2018b} D. Wang, L. Kong, P. Fan, H. Chen, S. Zhu, W. Liu, L. Cao, Y. Sun, S. Du, J. Schneeloch, R. Zhong, G. Gu, L. Fu, H. Ding, and H.-J. Gao, Science {\bf 362}, 333 (2018).

\bibitem{Marcus2019} A. Fornieri, A. M. Whiticar, F. Setiawan, E. Portol{\'e}s, A. C. Drachmann, A. Keselman, S. Gronin, C. Thomas, T. Wang, R. Kallaher, G. C. Gardner, E. Berg, M. J. Manfra, A. Stern, C. M. Marcus, and F. Nichele, Nature {\bf 569}, 89 (2019).

\bibitem{Molenkamp2019} H. Ren, F. Pientka, S. Hart, A. T. Pierce, M. Kosowsky, L. Lunczer, R. Schlereth, B. Scharf, E. M. Hankiewicz, L. W. Molenkamp, B. I. Halperin, and A. Yacoby, Nature {\bf 569}, 93 (2019).

\bibitem{Yazdani2019} B. J{\aa}ck, Y. Xie, J. Li, S. Jeon, B. A. Bernevig, and A. Yazdani, Science {\bf 364}, 1255 (2019).

\bibitem{SCZhang2009} X.-L. Qi, T. L. Hughes, S. Raghu, and S.-C. Zhang, Phys. Rev. Lett. {\bf 102}, 187001 (2009).

\bibitem{TeoKane2010} J. C. Y. Teo and C. L. Kane, Phys. Rev. B {\bf 82}, 115120 (2010).

\bibitem{Timm2010} A. P. Schnyder, P. M. R. Brydon, D. Manske, and C. Timm, Phys. Rev. B {\bf 82}, 184508 (2010).

\bibitem{Beenakker2011} C. W. J. Beenakker, J. P. Dahlhaus, M. Wimmer, and A. R. Akhmerov, Phys. Rev. B {\bf 83}, 085413 (2011).

\bibitem{KTLaw2012} C. L. M. Wong and K. T. Law, Phys. Rev. B {\bf 86}, 184516 (2012).

\bibitem{Nagaosa2012} S. Nakosai, Y. Tanaka, and N. Nagaosa, Phys. Rev. Lett. {\bf 108}, 147003 (2012).

\bibitem{KaneMele2013} F. Zhang, C. L. Kane, and E. J. Mele, Phys. Rev. Lett. {\bf 111}, 056402 (2013).

\bibitem{Berg2013} A. Keselman, L. Fu, A. Stern, and E. Berg, Phys. Rev. Lett. {\bf 111}, 116402 (2013).

\bibitem{Oreg2019} A. Haim and Y. Oreg, Physics Reports {\bf 825}, 1 (2019).

\bibitem{XJLiu2014} X.-J. Liu, C. L. M. Wong, and K. T. Law, Phys. Rev. X {\bf 4}, 021018 (2014).

\bibitem{XJLiu2016} P. Gao, Y.-P. He, and X.-J. Liu, Phys. Rev. B {\bf 94}, 224509 (2016).

\bibitem{Cooper2018} M. McGinley and N. R. Cooper, Phys. Rev. Lett. {\bf 121}, 090401 (2018).

\bibitem{localmixing1} K. W\"olms, A. Stern, and K. Flensberg, Phys. Rev. Lett. {\bf 113}, 246401 (2014).

\bibitem{localmixing2} K. W\"olms, A. Stern, and K. Flensberg, Phys. Rev. B {\bf 93}, 045417 (2016).

\bibitem{Knapp2020} C. Knapp, A. Chew, and J. Alicea, arXiv: 2006.10772.

\bibitem{spintri2003} A. P. Mackenzie and Y. Maeno, Rev. Mod. Phys. {\bf 75}, 657 (2003).

\bibitem{spintri2006} S. Das Sarma, C. Nayak, and S. Tewari, Phys. Rev. B {\bf 73}, 220502 (2006).

\bibitem{spintri2008} X. Dai, Z, Fang, Y. Zhou,  and F. -C. Zhang, Phys. Rev. Lett. {\bf 101}, 057008 (2008).

\bibitem{spintri2010} J. D. Strand, D. J. Bahr, D. J. Van Harlingen, J. P. Davis, W. J. Gannon, and W. P. Halperin, Science {\bf 328}, 1368 (2010).

\bibitem{spintri2010b} L. Fu and E. Berg, Phys. Rev. Lett. {\bf 105}, 097001 (2010).

\bibitem{spintri2019} S. Ran, C. Eckberg, Q. -P. Ding, Y. Furukawa, T. Metz, S. R. Saha, I.-L. Liu, M. Zic, H. Kim, J. Paglione, and N. P. Butch, Science {\bf 365}, 684 (2019).

\bibitem{ORL2014} X. -J. Liu, K. -T. Law, and T. -K. Ng, Phys. Rev. Lett. {\bf 112}, 086401 (2014).

\bibitem{ORL2016} Z. Wu, L. Zhang, W. Sun, X. -T. Xu, B. -Z. Wang, S. -C. Ji, Y. Deng, S. Chen, X. -J. Liu, and J. -W. Pan, Science {\bf 354}, 83 (2016).

\bibitem{ORL2019} Y.-H. Lu, B.-Z. Wang, and X.-J. Liu, arXiv:1911.07169; Science Bulletin {\bf 65}, 2080 (2020), doi: https://doi.org/10.1016/j.scib.2020.09.036

\bibitem{ORL2020} Z.-Y. Wang, X.-C. Cheng, B.-Z. Wang, J.-Y. Zhang, Y.-H. Lu, C.-R. Yi, S. Niu, Y. Deng, X.-J. Liu, S. Chen, and J.-W. Pan, arXiv:2004.02413.
    
\bibitem{Kotetes2013} P. Kotetes, New J. Phys. {\bf 15}, 105027 (2013).

\bibitem{GilbertBernevig2014} C. Fang, M. J. Gilbert, and B. A. Bernevig, Phys. Rev. Lett. {\bf 112}, 106401 (2014).

\bibitem{XJLiu2014C4} X.-J. Liu, J. J. He, and K. T. Law, Phys. Rev. B {\bf 90}, 235141 (2014).

\bibitem{Ryu2016review} C.-K. Chiu, J. C. Y. Teo, A. P. Schnyder, and S. Ryu, Rev. Mod. Phys. {\bf 88}, 035005 (2016).

\bibitem{Nagaosa2014} R. Wakatsuki, M. Ezawa, and N. Nagaosa, Phys. Rev. B {\bf 89}, 174514 (2014).

\bibitem{SM} See Supplemental Material for more details.

\bibitem{AliceaReview2012} J. Alicea, Rep. Prog. Phys. {\bf 75}, 076501 (2012).

\bibitem{ShenBook} S.-Q. Shen, {\it Topological Insulators}, Vol. 174 (Springer,
2012).



\bibitem{LG1} L. Allen, M. W. Beijersbergen, R. J. C. Spreeuw, and J. P. Woerdman, Phys. Rev. A {\bf 45}, 8185 (1992).

\bibitem{LG2} G. Juzeli\ifmmode \bar{u}\else \={u}\fi{}nas and P. \"Ohberg, Phys. Rev. Lett. {\bf 93}, 033602 (2004).

\end{thebibliography}

\begin{thebibliography}{99}

\bibitem[S1]{AliceaReview} J. Alicea, Rep. Prog. Phys. {\bf 75}, 076501 (2012).

\bibitem[S2]{WeylTheory2019} Y.-H. Lu, B.-Z. Wang, and X.-J. Liu, arXiv:1911.07169.

\bibitem[S3]{WeylExp2020} Z.-Y. Wang, X.-C. Cheng, B.-Z. Wang, J.-Y. Zhang, Y.-H. Lu, C.-R. Yi, S. Niu, Y. Deng, X.-J. Liu, S. Chen, and J.-W. Pan, arXiv:2004.02413.

\bibitem[S4]{Deutsch1998} I. H. Deutsch and P. S. Jessen, Phys. Rev. A {\bf 57}, 1972 (1998).

\bibitem[S5]{Mandel2003} O. Mandel, M. Greiner, A. Widera, T. Rom, T. W. H{\"a}nsch, and I. Bloch, Phys. Rev. Lett. {\bf 91}, 010407 (2003).

\bibitem[S6]{Wang2020} Y. Wang, L. Zhang, S. Niu, D. Yu, and X.-J. Liu, Phys. Rev. Lett. {\bf125}, 073204 (2020).

\bibitem[S7]{ShakingExp1}  N. Gemelke, E. Sarajlic, Y. Bidel, S. Hong, and S. Chu, Phys. Rev. Lett. {\bf 95}, 170404 (2005).

\bibitem[S8]{ShakingExp2} C. V. Parker, L.-C. Ha, and C. Chin, Nat. Phys. {\bf 9}, 769 (2013).

\end{thebibliography}

\setcounter{equation}{0} \setcounter{figure}{0}
\setcounter{table}{0}
\renewcommand{\theparagraph}{\bf}
\renewcommand{\thefigure}{S\arabic{figure}}
\renewcommand{\theequation}{S\arabic{equation}}

\onecolumngrid
\flushbottom
\newpage

\begin{center}\large
\textbf{Supplemental Material}
\end{center}

In this Supplemental Material, we provide details for the generic theory of unitary symmetry-protected non-Abelian statistics, the model Hamiltonian for TSC, the braiding matrices in different representations and the experimental realization.

\section{Unitary symmetry protection}
In this section, we show in general that in a unitary symmetry-protected TSC, the braiding of two vortices with each hosting multiple Majorana modes can be reduced into multiple independent sectors, each of which braids independently, leading to the symmetry-protected non-Abelian statistics. The braiding process is described by the evolution operator
\begin{eqnarray}
U(T)= \lim_{\Delta t\rightarrow0}\hat{T}\mathrm{e}^{-i\int_{-T/2}^{T/2}\mathrm{d}\tau H(\tau)}
= \mathrm{e}^{-iH(T/2)\Delta t}\mathrm{e}^{-iH(T/2-\Delta t)\Delta t}\cdots\mathrm{e}^{-iH(-T/2)\Delta t},
\end{eqnarray}
where $\hat{T}$ is the time-ordering operator. If the system keeps a unitary symmetry $\mathcal{R}$ throughout the braiding process, i.e. $[H(t),\mathcal{R}]=0$ at any time for $-T/2\leq t\leq T/2$, where $T$ is the total braiding time and $\mathcal{R}$ is the unitary symmetry that protects the topology, the effective Hamiltonian $H_E=iT^{-1}\text{ln} U(T)$ also satisfies the symmetry $\mathcal{R}$ since
\begin{eqnarray}
\mathcal{R}U(T)\mathcal{R}^{-1}
&=& \lim_{\Delta t\rightarrow0}\mathcal{R}\mathrm{e}^{-iH(T/2)\Delta t}\mathcal{R}^{-1}\mathcal{R}\mathrm{e}^{-iH(T/2-\Delta t)\Delta t}\mathcal{R}^{-1} \cdots\mathcal{R}\mathrm{e}^{-iH(-T/2)\Delta t}\mathcal{R}^{-1}\nonumber\\
&=& \lim_{\Delta t\rightarrow0}\mathrm{e}^{-iH(T/2)\Delta t}\mathrm{e}^{-iH(T/2-\Delta t)\Delta t}\cdots\mathrm{e}^{-iH(-T/2)\Delta t}\nonumber\\
&=& U(T),
\end{eqnarray}
so that $[U(T),\mathcal{R}]=[H_E,\mathcal{R}]=0$. Now we consider the case with each vortex binding $N$ Majorana modes $\tilde\gamma_j$ ($j=1,2,...,N$), between each pair of the Majorana modes, e.g. $\tilde\gamma_i$ and $\tilde\gamma_j$ there is a unitary sub-symmetry $\mathcal{R}_{ij}$ providing the protection. Note that since the braiding process is a unitary transformation which exchange the positions of two vortices, the effective Hamiltonian must be a
linear one in terms of Majorana operators. The effective Hamiltonian takes
the generic form $H_{E}=i\sum_{ij}\epsilon_{ij}\tilde{\gamma}_{i}^{L}\tilde{\gamma}_{j}^{R}$ since the local coupling is excluded by the protection of unitary symmetries. So in the Majorana bases, the effective Hamiltonian can be rewritten as
\begin{equation}
H_{E}=i\left(\begin{matrix}0 & H_{1}\\
-H_{1}^{\text{T}} & 0
\end{matrix}\right),
\end{equation}
where $H_{1}$ is a real matrix. The braiding operator $U=\exp(-iH_{E}T)=\sum\frac{1}{n!}(-iH_{E}T)^{n}$, in which
\begin{equation}
(-iH_{E})^{2n}=\left(
\begin{matrix}
-H_{1}H_{1}^{\text{T}} & 0\\
0 & -H_{1}^{\text{T}}H_{1}
\end{matrix}\right)^{n}; \quad
(-iH_{E})^{2n+1}=\left(
\begin{matrix}0 & (-H_{1}H_{1}^{\text{T}})^{n}H_{1}\\
(-H_{1}^{\text{T}}H_{1})^{n}(-H_{1}^{T}) & 0
\end{matrix}\right).
\end{equation}
The braiding process exchanges the left and right hand vortices, so the left and right hand side MZMs must exchange positions, and accordingly the diagonal elements in $U$ must vanish by definition. Then only
the off-diagonal blocks survive, giving
\begin{equation}
U=\left(\begin{matrix}0 & U_{1}\\
-U_{1}^{\text{T}} & 0
\end{matrix}\right).
\end{equation}
Since $U$ is unitary, one can verify that $U_{1}$ is also unitary. Therefore $U^2=-1$ by noting that $U_1^T=U_1^\dag$, which means each MZM accumulates a $\pi$ phase after full braiding. The phase after a single braiding can be fixed by
$$\gamma_{i}^{R}=\sum_{j}(-U_{1}^{\text{T}})_{ji}\tilde{\gamma}_{j}^{R},$$ so that
\begin{equation}
\gamma_i^{L}\to\gamma_i^{R}\quad\gamma_i^{R}\to-\gamma_i^{L}.
\end{equation}
This implies that under unitary symmetry protection, the MZMs can always be effectively reduced into several individual sectors, which are characterized by the braiding operator given by Eq.~(1) of the main text. Note that in general $\gamma_i^{L/R}$ can be a complicated superposition of $\tilde\gamma_j^{L/R}$s, depending on the details of the braiding process and the static Hamiltonian. However, during the braiding a MZM in one vortex can only effectively see one, rather than all MZMs in another vortex. This essentially renders the symmetry-protected non-Abelian statistics.

\section{Model Hamiltonian}

The tight-binding Hamiltonian for our model TSC is
\begin{eqnarray}\label{SC-SI}
H_{\text{TSC}}&=& -t_{0}\sum_{\langle\bm{i}\bm{j}\rangle}\left(c_{\bm{i}\uparrow}^{\dagger}c_{\bm{j}\uparrow}+c_{\bm{i}\downarrow}^{\dagger}c_{\bm{j}\downarrow}\right)-\mu'\sum n_{\bm{i}}
 +\left[\sum_{j_{x}}\Delta\left(c_{j_{x}\uparrow}^{\dagger}c_{j_{x}+1\downarrow}^{\dagger}-c_{j_{x}\uparrow}^{\dagger}c_{j_{x}-1\downarrow}^{\dagger}\right)+\text{h.c.}\right]\nonumber\\
 &&+\left[\sum_{j_{y}}i\Delta\left(c_{j_{y}\uparrow}^{\dagger}c_{j_{y}+1\downarrow}^{\dagger}-c_{j_{y}\uparrow}^{\dagger}c_{j_{y}-1\downarrow}^{\dagger}\right)+\text{h.c.}\right],
\end{eqnarray}
and the model is protected by unitary symmetries $\mathcal{S}_{1(2)}$, whose second-quantized forms are $\mathcal{S}_{1(2)}(c_\uparrow,c_\downarrow)^{\text{T}}\mathcal{S}_{1(2)}^\dagger=\sigma_{x(y)}(c_\uparrow,c_\downarrow)^{\text{T}}$. It's convenient to rewrite the Hamiltonian in $\bm{k}$-space, $H_{\text{TSC}}=1/2\sum_{\bm{p}}\Psi_{\bm{p}}^\dagger[\epsilon(\bm{p})\tau_z\otimes\sigma_0-2|\Delta|(\sin{p_x}\tau_y+\sin{p_y}\tau_x)\otimes\sigma_x]\Psi_{\bm{p}}$, where $\epsilon(\bm{p})=-2t(\cos{p_x}+\cos{p_y}-\mu')$ and the phase of $\Delta$ is absorbed into the definition of $\Psi_{\bm{p}}$. In the continuum limit, the single-particle BdG Hamiltonian is transformed into (\ref{Ham}) with $\mu=\mu'+4t_0$ and $m=1/2t_0$. The vortex in this system traps two MZMs in topological phase, and in the limit $m\to\infty$, they take the form
\begin{eqnarray}
\gamma_{a}&=&\frac{1}{\mathcal{N}}\int r\mathrm{d}r\mathrm{d}\theta f(r)[-\mathrm{e}^{i\theta/2}c_{\uparrow}(r,\theta)+\mathrm{e}^{i\theta/2}c_{\downarrow}(r,\theta)-\mathrm{e}^{-i\theta/2}c_{\uparrow}^{\dagger}(r,\theta)+\mathrm{e}^{-i\theta/2}c_{\downarrow}^{\dagger}(r,\theta)],\\
\gamma_{b}&=&\frac{i}{\mathcal{N}}\int r\mathrm{d}r\mathrm{d}\theta f(r)[\mathrm{e}^{i\theta/2}c_{\uparrow}(r,\theta)+\mathrm{e}^{i\theta/2}c_{\downarrow}(r,\theta)-\mathrm{e}^{-i\theta/2}c_{\uparrow}^{\dagger}(r,\theta)-\mathrm{e}^{-i\theta/2}c_{\downarrow}^{\dagger}(r,\theta)].
\end{eqnarray}
The MZMs will be transformed by the symmeties as $\mathcal{S}_1\gamma_a\mathcal{S}_1^\dagger=-\gamma_a$, $\mathcal{S}_1\gamma_b\mathcal{S}_1^\dagger=\gamma_b$ and $\mathcal{S}_2\gamma_a\mathcal{S}_2^\dagger=\gamma_b$, $\mathcal{S}_2\gamma_b\mathcal{S}_2^\dagger=\gamma_a$, which means the pair of MZMs are symmetry-protected.

\section{Braiding matrices}

The braiding operator $\mathcal{B}_{ij}=\mathrm{e}^{-\frac{\pi}{4}\gamma_{ia}\gamma_{ja}}\mathrm{e}^{-\frac{\pi}{4}\gamma_{ib}\gamma_{jb}}$
is particle-number conserved in the $\eta$-representation thus the braiding matrices are block-diagonal in particle-number conserved
subspaces. For four vortices, the state of system can be generally written as
$|n_{1-}n_{1+}\rangle_{\eta}|n_{2-}n_{2+}\rangle$,
where $n_{1(2)\pm}$ denotes the particle number of $\eta_{1(2)\pm}$.
In the one-particle subspace $\{|10\rangle_{\eta}|00\rangle_{\eta},|01\rangle_{\eta}|00\rangle_{\eta},|00\rangle_{\eta}|10\rangle_{\eta},|00\rangle_{\eta}|01\rangle_{\eta}\}$,
or three-particle subspace \{$|10\rangle_{\eta}|11\rangle_{\eta}$,
$-|01\rangle_{\eta}|11\rangle_{\eta}$, $|11\rangle_{\eta}|10\rangle$,
$-|11\rangle_{\eta}|01\rangle_{\eta}$\}, the braiding operators are
\begin{eqnarray}
&&B_{12}=\begin{pmatrix}
i\sin\psi_{12} & -\cos\psi_{12} & 0 & 0\\
\cos\psi_{12} & -i\sin\psi_{12} & 0 & 0\\
0 & 0 & 1 & 0\\
0 & 0 & 0 & 1\end{pmatrix},
B_{34}=\begin{pmatrix}
1 & 0 & 0 & 0\\
0 & 1 & 0 & 0\\
0 & 0 & i\sin\psi_{34} & -\cos\psi_{34}\\
0 & 0 & \cos\psi_{34} & -i\sin\psi_{34}\end{pmatrix},\nonumber\\
&&\hspace{60pt}B_{23}=\frac{1}{2}\begin{pmatrix}
1 & 1 & \mathrm{e}^{-i\psi_{12}} & \mathrm{e}^{-i\psi_{12}}\\
1 & 1 & -\mathrm{e}^{-i\psi_{12}} & -\mathrm{e}^{-i\psi_{12}}\\
-\mathrm{e}^{i\psi_{12}} & \mathrm{e}^{i\psi_{12}} & 1 & -1\\
-\mathrm{e}^{i\psi_{12}} & \mathrm{e}^{i\psi_{12}} & -1 & 1
\end{pmatrix}.
\end{eqnarray}
The operator $\mathcal{B}_{12(34)}$ only affects $\eta_{\alpha(\beta)\pm}$ while $\mathcal{B}_{23}$ combines all four states. In the two-particle subspace \{$|11\rangle_{\eta}|00\rangle_{\eta}$,
$|00\rangle_{\eta}|11\rangle_{\eta}$, $|10\rangle_{\eta}|10\rangle_{\eta}$,
$|10\rangle_{\eta}|01\rangle_{\eta}$, $|01\rangle_{\eta}|10\rangle_{\eta}$,
$|01\rangle_{\eta}|01\rangle_{\eta}$\}, the braiding operators are
\begin{eqnarray}
&&B_{12}=\begin{pmatrix}
1 & 0 & 0 & 0 & 0 & 0\\
0 & 1 & 0 & 0 & 0 & 0\\
0 & 0 & i\sin\psi_{12} & 0 & -\cos\psi_{12} & 0\\
0 & 0 & 0 & i\sin\psi_{12} & 0 & -\cos\psi_{12}\\
0 & 0 & \cos\psi_{12} & 0 & -i\sin\psi_{12} & 0\\
0 & 0 & 0 & \cos\psi_{12} & 0 & -i\sin\psi_{12}
\end{pmatrix},
B_{34}=\begin{pmatrix}
1 & 0 & 0 & 0 & 0 & 0\\
0 & 1 & 0 & 0 & 0 & 0\\
0 & 0 & i\sin\psi_{34} & -\cos\psi_{34} & 0 & 0\\
0 & 0 & \cos\psi_{34} & -i\sin\psi_{34} & 0 & 0\\
0 & 0 & 0 & 0 & i\sin\psi_{34} & -\cos\psi_{34}\\
0 & 0 & 0 & 0 & \cos\psi_{34} & -i\sin\psi_{34}
\end{pmatrix},\nonumber\\
&&\hspace{100pt}B_{23}=\frac{1}{2}\begin{pmatrix}
0 & 0 & -\mathrm{e}^{-i\psi_{12}} & -\mathrm{e}^{-i\psi_{12}} & -\mathrm{e}^{-i\psi_{12}} & -\mathrm{e}^{-i\psi_{12}}\\
0 & 0 & \mathrm{e}^{i\psi_{12}} & -\mathrm{e}^{i\psi_{12}} & -\mathrm{e}^{i\psi_{12}} & \mathrm{e}^{i\psi_{12}}\\
\mathrm{e}^{i\psi_{12}} & -\mathrm{e}^{-i\psi_{12}} & 1 & 0 & 0 & -1\\
\mathrm{e}^{i\psi_{12}} & \mathrm{e}^{-i\psi_{12}} & 0 & 1 & -1 & 0\\
\mathrm{e}^{i\psi_{12}} & \mathrm{e}^{-i\psi_{12}} & 0 & -1 & 1 & 0\\
\mathrm{e}^{i\psi_{12}} & -\mathrm{e}^{-i\psi_{12}} & -1 & 0 & 0 & 1
\end{pmatrix}.
\end{eqnarray}
The braiding operators act trivially on the vaccum state and four-particle
state $|11\rangle_{\eta}|11\rangle_{\eta}$.

In the $f$-representation, braiding operator $\mathcal{B}_{ij}=\mathrm{e}^{-\frac{\pi}{4}\gamma_{ia}\gamma_{ja}}\mathrm{e}^{-\frac{\pi}{4}\gamma_{ib}\gamma_{jb}}$
is parity-conserved in sectors $a$ and $b$, and the braiding matrices
are direct product of matrices in each sector. The particle-number
conservation of $\eta$ and the parity conservation of $f$ are not
contradictory since $f$ is generally the linear combination of $\eta$
and $\eta^{\dagger}$, for example, $f_{1a}=(\eta_{1}+i\eta_{2}+\eta_{1}^{\dagger}+i\eta_{2}^{\dagger})/2$,
and therefore the vaccum states $|0\rangle_{\eta}$ and $|0\rangle_{f}$
are different. For four vortices, the state of system can be generally written as
$|n_{1a}n_{1b}\rangle_{f}|n_{2a}n_{2b}\rangle_{f}$, where $n_{ia(b)}$
denotes the particle-number of $f_{ia(b)}$. The whole Fock space can be divided into even and odd parity parts of $a(b)$ sectors
\begin{equation}
\begin{array}{l@{\ }l@{\ }l@{\ }}
a\,\text{Even}\;&b\,\text{Even}&:\{|00\rangle_{f}|00\rangle_{f}, |01\rangle_{f}|01\rangle_{f}, |10\rangle_{f}|10\rangle_{f}, |11\rangle_{f}|11\rangle_{f}\},\\
a\,\text{Even}\;&b\,\text{Odd}&:\{|01\rangle_{f}|00\rangle_{f}, |00\rangle_{f}|01\rangle_{f}, -|11\rangle_{f}|10\rangle_{f}, |10\rangle_{f}|11\rangle_{f}\},\\
a\,\text{Odd}\;&b\,\text{Even}&:\{|10\rangle_{f}|00\rangle_{f}, |11\rangle_{f}|01\rangle_{f}, |00\rangle_{f}|10\rangle_{f}, -|01\rangle_{f}|11\rangle_{f}\},\\
a\,\text{Odd}\;&b\,\text{Odd}&:\{|11\rangle_{f}|00\rangle_{f}, |10\rangle|01\rangle_{f}, -|01\rangle_{f}|10\rangle_{f}, -|00\rangle_{f}|11\rangle_{f}\}.
\end{array}
\end{equation}
Braiding matrices in a single sector can be written in even and odd parity subspaces~\cite{AliceaReview}
\begin{equation}
\begin{array}{cc}
B_{12}=\begin{cases}
\begin{pmatrix}
\mathrm{e}^{-i\frac{\pi}{4}} & 0\\
0 & \mathrm{e}^{i\frac{\pi}{4}}
\end{pmatrix}\vspace*{5pt} & \text{for even parity;}\\
\begin{pmatrix}
\mathrm{e}^{i\frac{\pi}{4}} & 0\\
0 & \mathrm{e}^{-i\frac{\pi}{4}}
\end{pmatrix} & \text{for odd parity,}
\end{cases} &
\begin{array}{c}
B_{34}=\begin{pmatrix}
\mathrm{e}^{-i\frac{\pi}{4}} & 0\\
0 & \mathrm{e}^{i\frac{\pi}{4}}
\end{pmatrix}\vspace*{5pt},\\
B_{23}=\frac{1}{\sqrt{2}}\begin{pmatrix}
1 & -i\\
-i & 1
\end{pmatrix}.
\end{array}
\end{array}
\end{equation}

\begin{figure}
\includegraphics[width=0.65\textwidth]{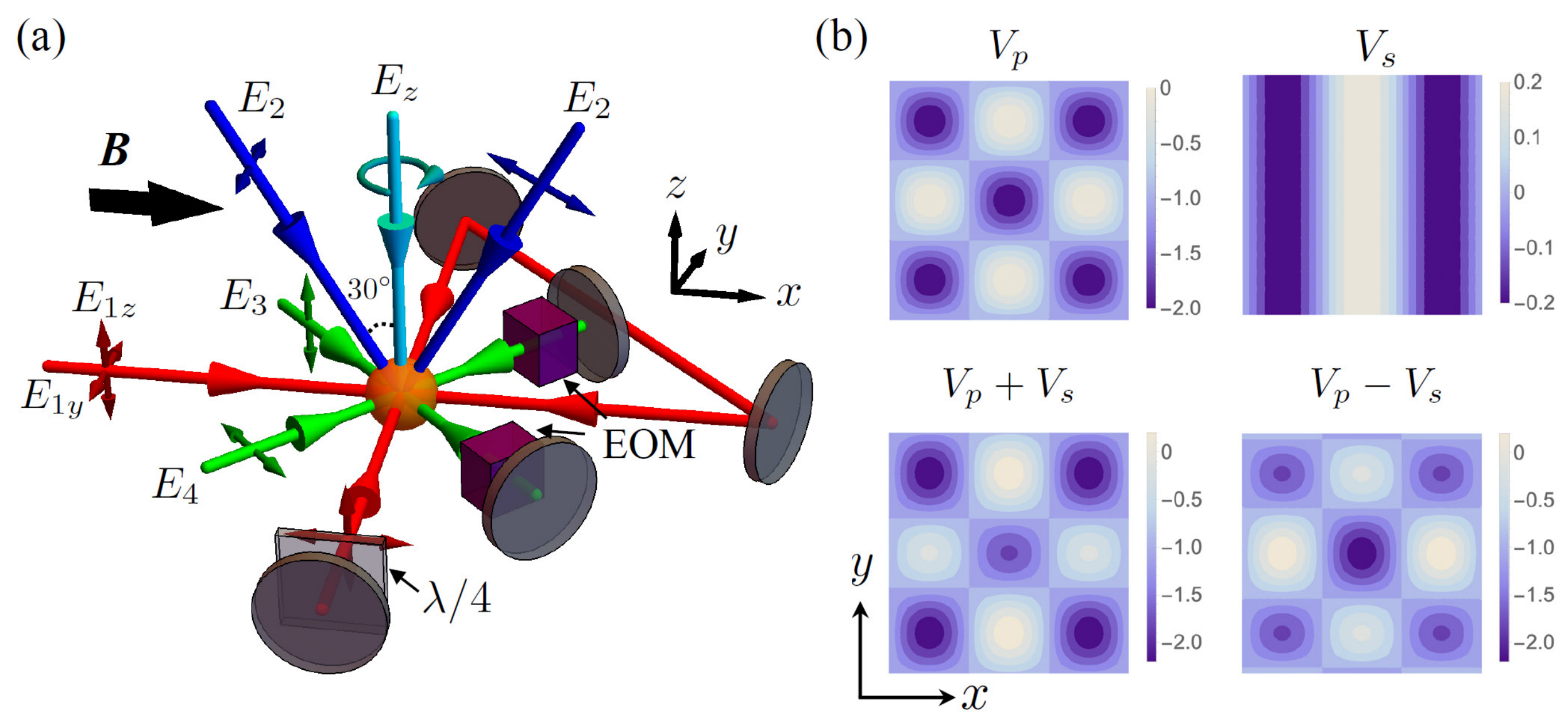}
\caption{Experimental realization in an optical Raman lattice. (a) Experimental setup.
With the bias magnetic field ${\bm B}$ applied in the $x$ direction, the light components $E_{1y,1z}$ (red)
form a spin-independent checkerboard lattice, and generate periodic Raman coupling potentials together with the circularly polarized beam $E_z$ incident along the $z$-axis. The light $E_z$ is set to be
a Laguerre-Gaussian (LG) beam carrying optical vortices, which imprints a spatially-dependent phase on spin-flipped hoppings.
Two beam components of $E_2$ (blue) with mutually perpendicular polarization is used to produce a spin-dependent lattice in the $x$ direction, which provides a staggered Zeeman energy splitting.
Two beams $E_{3,4}$ are applied along the $x+y$ and $x-y$ directions respectively to form a 2D shaking lattice by sinusoidally modulating the phases of retro-reflected components via electro-optic modulators (EOMs). (b) The contour plot of the lattice potentials $V_p$ and $V_s$ (upper panel). Their combination produces a staggered on-site energy difference for the two spin states (lower panel).
Here $V_1=-0.4V_0$.
}\label{figS1}
\end{figure}

\section{Experimental realization}

We propose an experimental setup based on the optical Raman lattice scheme which was used to realize Weyl semimetal in 3D~\cite{WeylTheory2019,WeylExp2020}, as sketched in Fig.~\ref{figS1}.
We aim to realize the Hamiltonian
\begin{align}\label{Htot_S}
\begin{split}
H=&H_{0}+V_s({\bm r})\sigma_z+V_{t}({\bm r},t)\otimes{\bf 1},\\
H_0=&\left[\frac{{\bm k}^2}{2m}+V_{p}({\bm r})\right]\otimes{\bf 1}+{\cal M}({\bm r})|\!\uparrow\rangle\langle\downarrow\!|+{\cal M}^*({\bm r})|\!\downarrow\rangle\langle\uparrow\!|
\end{split}
\end{align}
where $V_{p}({\bm r})$ is the primary checkerboard  lattice, ${\cal M}({\bm r})$ denotes the Raman coupling potential with a spatially-dependent phase,
$V_s({\bm r})$ is a spin-dependent lattice, which provides a staggered Zeeman splitting at each lattice site and thus suppresses the on-site spin flipping,
and $V_t({\bm r},t)$ is a two-dimensional (2D) shaking lattice, which is applied to induce and control the spin-conserved nearest-neighbor hopping.


\subsection{Primary checkerboard lattice and Raman coupling potential}

As shown in Fig.~\ref{figS1}(a), the beams $E_{1y}$ of frequency $\omega_{1y}$ with $\hat{y}$-polarization and $E_{1z}$ of frequency $\omega_{1z}$ ($\omega_{1z}\neq\omega_{1y}$) with $\hat{z}$-polarization together form a 2D standing waves ${\bm E}_1={\bm E}_{1xy}+{\bm E}_{1z}$ in the $x$-$y$ plane via mirror reflection, where
\begin{align}
\begin{split}
{\bm E}_{1xy}&=E_{1y}(i\hat{y}e^{ik_0x}-i\hat{y}e^{-ik_0x}+i\hat{x}e^{-ik_0y}-i\hat{x}e^{ik_0y})e^{-i\omega_{1y}t},\\
&=2E_{1y}(\hat{x}\sin k_0y-\hat{y}\sin k_0x)e^{-i\omega_{1y}t},\\
{\bm E}_{1z}&=E_{1z}\hat{z}(e^{ik_0x}+e^{-ik_0x}+e^{ik_0y}+e^{-ik_0y})e^{-i\omega_{1z}t},\\
&=2E_{1z}\hat{z}(\cos k_0x+ \cos k_0y)e^{-i\omega_{1z}t}.
\end{split}
\end{align}
Note that the relative phase between the $z$- and $x$-polarized components acquires a $\pi$ phase after passing through the quarter-wave plate ($\lambda/4$) twice.
When $E_{1y}=E_{1z}=E_0$, the standing wave ${\bm E}_1$ leads to a checkerboard lattice $V_{p}({\bm r})\propto|{\bm E}_1|^2$~\cite{WeylTheory2019}, which takes the form
\begin{align}\label{Vp_S}
V_{p}({\bm r})=V_0\cos^2(k'_0x')+V_0\cos^2(k'_0y'),
\end{align}
where $V_0\propto E_0^2$, $k'_0=k_0/\sqrt{2}$, $x'=(x-y)/\sqrt{2}$, and $y'=(x+y)/\sqrt{2}$.

We further apply a circularly polarized Laguerre-Gaussian (LG) beam of frequency $\omega_{z}$ in the $z$ direction, which carries an orbital angular momentum $l$.
The light field can be written as
\begin{align}
{\bm E}_{z}=E_{z}\frac{\hat{x}+i\hat{y}}{\sqrt{2}}e^{il\theta(\bm r)}e^{i(k_0z-\omega_{z}t)},
\end{align}
where $\theta$ is the azimuthal angle.
If $\omega_{z}=\omega_{1y}+\omega_{\rm ZS}$ with $\omega_{\rm ZS}$ being the Zeeman splitting of the two spins,
the Raman potential ${\cal M}({\bm r})$ can be generated through the double-$\Lambda$-type coupling configuration via the standing wave ${\bm E}_{1xy}$ and the running wave ${\bm E}_{z}$, which reads
\begin{align}
{\cal M}({\bm r})=M_0e^{-\ui l\theta({\bm r})}(\sin k'_0x\cos k'_0y'-\ui\cos k'_0x\sin k'_0y'),
\end{align}
with $M_0\propto E_0E_z$.

\subsection{Spin-dependent lattice}

Two lights with mutually perpendicular polarization are applied in the $x$-$z$ plane to generate a spin-dependent lattice in the $x$ direction,
which are incident from the positive and negative $x$-axis, respectively, with the angles to the z-axis being both $30^\circ$ [see Fig.~\ref{figS1}(a)].
The light field reads
\begin{align}
{\bm E}_2=E_2(\hat{y}e^{ik_0(x+\sqrt{3}z)/2}+(\hat{z}-\sqrt{3}\hat{x})e^{ik_0(-x+\sqrt{3}z)/2}/2)e^{-i\omega_2t}.
\end{align}
The spin-dependent lattice is the vector light shift $V_s({\bm r})\propto i({\bm E}^*_2\times{\bm E}_2)\cdot\hat{x}$~\cite{Deutsch1998,Mandel2003}, thus taking the form
\begin{align}\label{Vs_S}
V_s({\bm r})=\frac{V_1}{2}\cos(k_0x),
\end{align}
with $V_1\propto E_2^2$. As shown in Fig.~\ref{figS1}(b), the combination of the primary lattice (\ref{Vp_S}) plus the spin-dependent lattice (\ref{Vs_S}) indeed provides a staggered on-site Zeeman term.
Note that for a realistic setup, one need to choose proper hyperfine states to construct the spin-$1/2$ system such as the ${\bm E}_2$-induced spin-independent part can be neglected.
For example, one can choose $|\!\uparrow\rangle=|F=7/2, m_F=+7/2\rangle$ and $|\!\downarrow\rangle=|9/2, +9/2\rangle$ for $^{40}$K atoms~\cite{Wang2020}.

%

\subsection{Shaking lattice}

The shaking lattice is formed by two beams of wavelength $\lambda_s=2\pi/k_0'$  applied in the $x'$ and $y'$ directions, respectively,
with the phases of their retro-reflected components being sinusoidally modulated~\cite{ShakingExp1,ShakingExp2}.
The light fields are
\begin{align}
\begin{split}
{\bm E}_3=&E_3\hat{z}\left[e^{i k'_0x'}+e^{-i(k'_0x'+\varphi_{x'})}\right]e^{-i\omega_st}=2E_3\hat{z}e^{-i\varphi_{x'}/2}\cos\left(k'_0x'+\frac{\varphi_{x'}}{2}\right)e^{-i\omega_st},\\
{\bm E}_4=&E_4\frac{\hat{x}-\hat{y}}{\sqrt{2}}\left[e^{i k'_0y'}+e^{-i(k'_0y'+\varphi_{y'})}\right]e^{-i\omega_st}=\sqrt{2}E_4(\hat{x}-\hat{y})e^{-i\varphi_{y'}/2}\cos\left(k'_0y'+\frac{\varphi_{y'}}{2}\right)e^{-i\omega_st},
\end{split}
\end{align}
Hence, the shaking lattice potential reads
\begin{align}
V_t({\bm r},t)=V_2\cos^2\left(k'_0x'+\frac{\varphi_{x'}(t)}{2}\right)+V_2\cos^2\left(k'_0y'+\frac{\varphi_{y'}(t)}{2}\right),
\end{align}
where $\varphi_{x'}(t)=f\cos(\omega_{\rm mod} t)$ and $\varphi_{y'}(t)=f\cos(\omega_{\rm mod} t+\phi_0)$ with $f$ being the shaking amplitude and $\omega_{\rm mod}\simeq V_1$.

\subsection{Tight-binding model}

In this subsection, we consider the tight-binding limit of the total Hamiltonian (\ref{Htot_S}).
We denote $\Phi_s^\sigma({\bm r})$ as the Wannier function of spin $\sigma=\uparrow\downarrow$ for $s$-bands of $V_{p}({\bm r}){\bf 1}+V_s({\bm r})\sigma_z$.
Without loss of generality, we suppose $V_0<0$.
The spin-flipped hopping is
\begin{align}
t_{\rm SO}=M_0\int \ud{\bm r} \Phi^\downarrow_s(x',y')\sin k'_0x\cos k'_0y'\Phi^\uparrow_s(x'-a,y'),
\end{align}
where $a$ denotes the lattice constant.
Due to the identity
\begin{align}
e^{\pm iz\cos\phi}&=\sum_{n=-\infty}^\infty (\pm i)^nJ_n(z)e^{in\phi}=J_0(z)+2\sum_{n=1}^\infty (\pm i)^nJ_n(z)\cos(n\phi),
\end{align}
where $J_n$ are the Bessel functions,
the shaking lattice potential can be written as $V_{t}({\rm r},t)\simeq V^{(0)}_{t}({\rm r})+V^{(1)}_{t}({\rm r},t)$, where
\begin{align}
\begin{split}
V^{(0)}_{t}({\rm r})&= V_2J_0(f)\left[\cos^2(k'_0x')+\cos^2(k'_0y')\right],\\
V^{(1)}_{t}({\rm r},t)&=-V_2J_1(f)\left[\sin(2k'_0x')\cos(\omega t)+\sin(2k'_0y')\cos(\omega t+\phi_0)\right].
\end{split}
\end{align}
The shaking-induced spin-conserved hopping can be written as
\begin{align}
t_0=-\frac{V_2J_1(f)}{2}\int \ud{\bm r} \Phi^\sigma_s(x',y')\sin(2k'_0x')\Phi_s^\sigma(x'-a,y').
\end{align}
With these ingredients the tight-binding model of QAHI is realized and given by Eq.~\eqref{SC-SI} after particle hole transformation as introduced in main text.

\subsection{Numerical Results}
\begin{figure}
\includegraphics[width=0.65\textwidth]{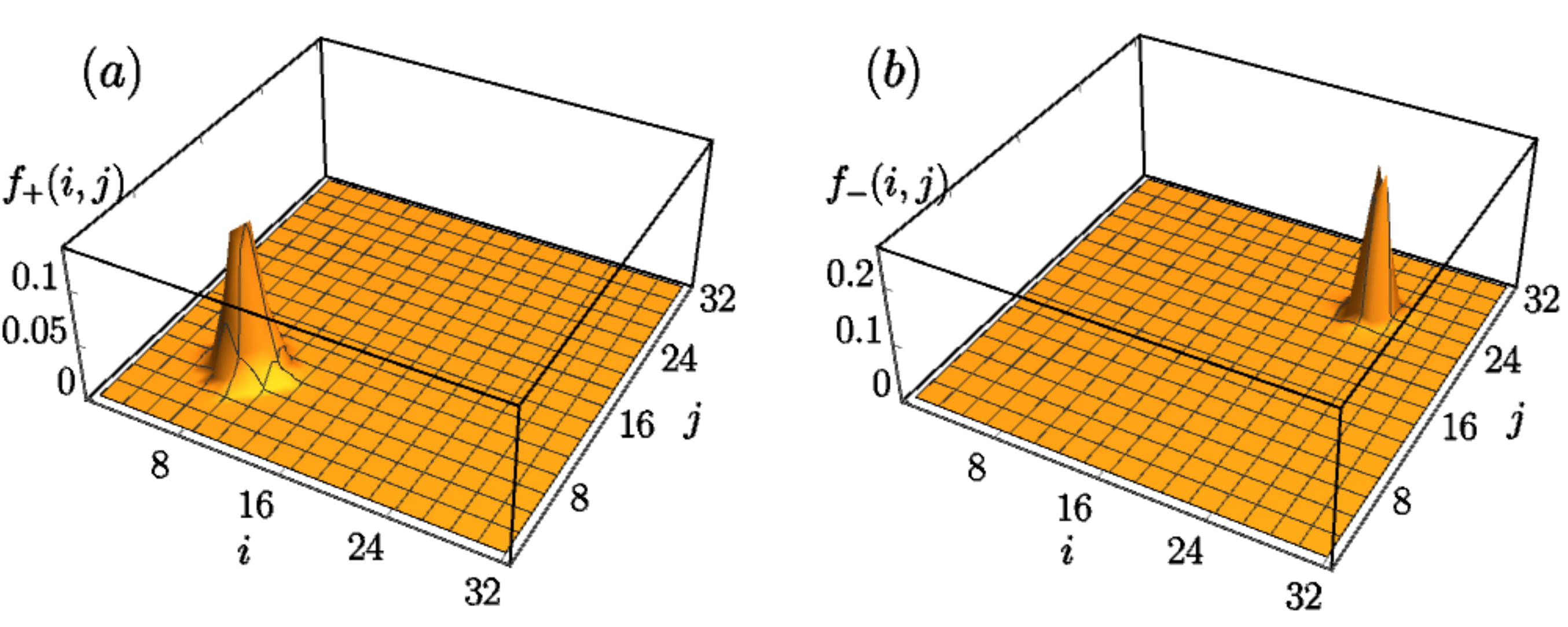}
\caption{$f_{\pm}(i,j)$ of two zero-modes which are localized at the vortex or anti-vortex respectively. The vortex is located at the barrier between $(8,8)$ and $(9,9)$, and the anti-vortex is located at the barrier between $(24,24)$ and $(25,25)$.
}\label{numwf}
\end{figure}

In the numerical simulation, we apply two LG beams that carry opposite angular momentum to eliminate the winding of Raman potential at the boundary. We confirmed that there are two almost degenerate eigenstates localized at the centers of the LG beams which energies are at the center of the gap opened by spin-orbit coupling. The remaining non-degeneracy is the result of two reasons - (i) There are residue effective $m_x$ or $m_y$ that are not canceled by the onsite staggered $m_z$ term, and (ii) the strength of $t_\text{SO}$ is spin and site dependent since the size of the barrier that are overcome by the Raman coupling has staggered pattern. The split of energy is about $1/57$ of the size of the gap with $V_0=5E_r, V_1=E_r, m_z=0.083E_r, V_2=5.5E_r$, $M_0=1.25E_r$ and $\omega_\text{mod}=0.434E_r$. Let $\psi_{\pm}$ be the wavefunctions of two zero modes and $f_{\pm}(i,j) = \int_{(x,y) \sim (i, j)} \left|(\psi_{\pm}(x,y)\right|^2 dx dy $, where the integration is done in the region of the site $(i,j)$, be the site-wise probability of the states. Fig.~\ref{numwf}(a-b) show $f_{\pm}(i,j)$ and the zero modes are localized at the centers of the vortex or anti-vortex.

\end{document}